\documentclass[prb,showpacs,preprintnumbers,amssymb,eqsecnum,twocolumn]{revtex4}
\usepackage{graphicx}
\usepackage{dcolumn}
\usepackage{bm}
\usepackage{amsmath, amssymb}
\usepackage{epstopdf}
\usepackage{epsfig}
\def\be{\begin{equation}}
\def\ee{\end{equation}}
\def\bea{\begin{eqnarray}}
\def\eea{\end{eqnarray}}

\begin{document}

\title{The Generalized DMPK equation revisited: A systematic derivation}
\author{Andrew Douglas$^a$, Peter Marko\v{s}$^b$ and K. A. Muttalib$^c$}
\affiliation{$^a$Department of Physics, University of North Florida, 1 UNF Dr.,
Jacksonville, FL 32224-7699}
\affiliation{$^b$Department of Physics, INPE, Faculty of Electrical Engineering and Information Technology, Slovak University of Technology, 81219 Bratislava, Slovakia}
\affiliation{$^c$Department of Physics, University of Florida, P.O. Box 118440,
Gainesville, FL 32611-8440}
\begin{abstract}
The Generalized Dorokov-Mello-Pereyra-Kumar (DMPK) equation has recently been used to obtain a family of very broad and highly asymmetric conductance distributions for three dimensional disordered conductors. However, there are two major criticisms of the derivation of the Generalized DMPK equation: (1) certain eigenvector correlations were neglected based on qualitative arguments that can not be valid for all disorder, and  (2) the repulsion between two closely spaced eigenvalues were not rigorously governed by symmetry considerations. In this work we show that it is possible to address both criticisms by including the eigenvalue and eigenvector correlations in a systematic and controlled way. It turns out that the added correlations determine the evolution of the Jacobian, without affecting the evaluation of the conductance distributions. They also guarantee the symmetry requirements. In addition, we obtain an exact relationship between the eigenvectors and the Lyapunov exponents leading to a sum rule for the latter at all disorder.

\end{abstract}


\maketitle

\section{Introduction}

A highly non-trivial, non-Gaussian distribution $P(g)$ of conductances $g$ for disordered three dimensional (3d) conductors in the large disorder limit has recently been predicted in Refs.~[\onlinecite{mmw,dm-10}]. The predictions include, among others, a strong deviation from the expected log-normal distribution in the deeply insulating limit in agreement with numerical results obtained from tight binding Anderson model  \cite{markos,soukoulis}, variance that grows with disorder  consistent with numerical simulations \cite{somoza}, and an asymmetry that changes sign near the metal-insulator transition as the disorder is decreased from the deeply insulating limit. The distribution was obtained using the so called Generalized DMPK equation \cite{mg}. While the original Dorokov-Mello-Pereyra-Kumar (DMPK) equation \cite{dmpk} has been shown to be valid for quasi one dimensional (1d) systems \cite{frahm} where there is no Anderson metal-insulator transition, the Generalized DMPK equation has been claimed to be valid beyond quasi 1d in the large disorder regime. Since the full distribution $P(g)$ in 3d at strong disorder is beyond the scope of the conventional field theory framework \cite{altshuler,shapiro}, and since a broad and highly asymmetric distribution can have interesting consequences for the Anderson transition, it is important to critically evaluate the validity of the Generalized DMPK equation. 

For a given $N$-channel conductor of length $L_z$ and cross section $L^2$, the $2N \times 2N$ transfer matrix $M$ relates the wavefunction on the left of the sample to that on the right via: $\Psi_R = M\Psi_L$.  For the case of time reversal and spin rotation symmetry (orthogonal ensemble), $M$ can be parameterized by the $N \times N$ unitary matrices $u$ and $\upsilon$, and $N \times N$ diagonal matrix $\lambda$:  
\bea
M &=&
\left(
\begin{array}{cc}
  u &   0   \\
 0 &   u^*    
\end{array}
\right)
\left(
\begin{array}{cc}
  \sqrt{1+\lambda} &   \sqrt{\lambda}  \\
  \sqrt{\lambda} &    \sqrt{1+\lambda}     
\end{array}
\right)
\left(
\begin{array}{cc}
 v & 0    \\
 0 & v^*  
\end{array}
\right)
\label{M}
\eea 
We will loosely refer to $\lambda$ and $u$, $v$ as the eigenvalues and the eigenvectors since they turn out to be the  eigenvalues and eigenvectors of a certain combination of $MM^{\dag}$, as will be shown later.
The $|u_{ab}|^2$ and $|\upsilon_{ab}|^2$ can be thought of as the fraction of current in channel $a$ that enters into channel $b$ from left to right and from right to left, respectively, and $1/(1+\lambda_b)$  is channel $b$'s contribution to the conductance. $M$ is a random matrix, and its probability distribution can be written as
\bea 
p(M)d\mu(M) = p(\lambda,u,\upsilon)d\mu(\lambda)d\mu(u)d\mu(\upsilon).
\eea
The measure $d\mu(\lambda)$ can be written in terms of the so-called Jacobian $J(\lambda)$:
\bea
d\mu(\lambda) = J(\lambda)d^N\lambda; \;\;\;\;\;\; J(\lambda)\equiv 
\prod_{a \neq b}^N|\lambda_a - \lambda_b|^{\beta_{}}
\label{Jac}
\eea
where the orthogonal symmetry discussed here requires \cite{mehta} $\beta=1$.
For future reference we'll define related functions:
\bea
&\;&p(\lambda,\upsilon) \equiv \int d\mu(u)p(\lambda,u,\upsilon) , \cr
&\;&p(\lambda) \equiv \int d\mu(u)d\mu(\upsilon) p(\lambda,u,\upsilon), \cr
&\;&P(\lambda) \equiv p(\lambda)J(\lambda).
\label{Pdef}
\eea

The last one is the probability distribution of the eigenvalues $\lambda_i$ ($i=1,2, \cdots N$).  One can obtain an 
evolution equation for $p(\lambda)$ in the following manner.  We add a small conductor of length $\delta L_z$ described by a transfer matrix $M_{\delta L_z}\equiv M'$ to a conductor of length $L_z$ described by $M_{L_z}\equiv M''$.  Then  taking advantage of the fact that the transfer matrices are multiplicative ($M_{L_z+\delta L_z}\equiv M=M''M'$), an integral equation for the evolution of $p(M)$ with length $L_z$ can be written down: 
\be
p_{L_z+\delta L_z} (M) = \int d\mu(M')p_{\delta L_z}(M')p_{L_z}(M'').
\label{start}
\ee
Integrating both sides over $d\mu(u)$ and $d\mu(\upsilon)$, we can write the evolution equation for $p(\lambda)$ as
\be
p_{L_z+\delta L_z} (\lambda) = \int d\mu(u)d\mu(\upsilon)\langle p_{L_z}(\lambda'',u'',\upsilon'')\rangle_{M'}
\label{pint}
\ee
where the angular bracket $\langle \cdots \rangle_{M'} $ indicates an averaging over $M'$.  When the sample width is much smaller than the localization length $\xi$ $(L << \xi)$, and the length is much larger than the width $(L_z >> L)$, the matrices $u$ and $v$ are isotropically distributed \cite{dmpk}. In that case, $p(\lambda,u,\upsilon)$ only depends on $\lambda$.
To convert the integral equation to a differential equation, $p(\lambda'')$ is expanded to second order in a Taylor series about $\lambda'' = \lambda$, and the average over $M'$ and integrals over $d\mu(u)$ and $d\mu(\upsilon)$ can be done.  The resulting evolution equation for $P(\lambda)$ (DMPK equation) is  
\bea
\frac{\partial P(\lambda)}{\partial t} &=& \frac{2}{N+1} 
\sum_{a=1}^N \frac{\partial}{\partial \lambda_a}\lambda_a(1+\lambda_a) \cr
&\;&\times \left[\frac{\partial}{\partial \lambda_a}-\sum_{b \neq a} \frac{1}{\lambda_a-\lambda_b} \right] P(\lambda)
\label{dmpk}
\eea
where $t \equiv L_z/l$, and $l$ is the mean free path.  One can obtain the conductance distribution from $P(\lambda)$ via the Landauer formula  \cite{mu-wo} 
 \be 
  \mathcal{P}(g)\propto \int\prod_{i=1}^N d^N\lambda P(\lambda)\delta
 \left (g-\sum_i \frac{1}{1+\lambda_i}\right ).
  \label{Pofg}
  \ee 
The $\mathcal{P}(g)$ obtained in this way has been shown to be valid for quasi 1d systems at all disorder, with surprising features near the metal-insulator crossover regime \cite{mu-wo,mwgg,gmw,mwg,martin}.

The first attempt towards relaxing the isotropy approximation was made in Ref.~[\onlinecite{mk}].
It was assumed that $p(\lambda,u,\upsilon)$ depended only on $\lambda$, but $u$ and $\upsilon$ were not isotropically distributed however.  The $p(\lambda'')$ was expanded out to second order about $\lambda'' = \lambda$.  The  average over $M'$ and integral over $d\mu(u)$ could be done and the integral over $d\mu(\upsilon)$ was parameterized via arbitrary constants $\mu_1$ and $\mu_2$.  Requiring that the resulting equation conserve probability necessiated a renormalization of the Jacobian from $J$ to $J^\gamma$ so that $P(\lambda)$ was redefined as
\be
\bar{P}(\lambda)= \prod_{a \neq b}^N|\lambda_a - \lambda_b|^\gamma p(\lambda).
\ee
It also linked $\mu_1$ and $\mu_2$ together leaving their ratio $\gamma \equiv \mu_1 / \mu_2$ as the only true degree of freedom, resulting in a one-parameter generalization of the DMPK equation:
\bea
\frac{\partial \bar{P}(\lambda)}{\partial t} &=& \frac{2\mu_2}{N+1} 
\sum_{a=1}^N \frac{\partial}{\partial \lambda_a}\lambda_a(1+\lambda_a) \cr
&\;&\times \left[\frac{\partial}{\partial \lambda_a}-\sum_{b \neq a} \frac{\gamma}{\lambda_a-\lambda_b} \right] \bar{P}(\lambda).
\label{FP}
\eea
The parameter $\mu_2$ just renormalizes the mean free path, while the presence of $\gamma$ evinces a disorder-dependent repulsion between the eigenvalues.  It was demonstrated in Ref. [\onlinecite{mk}] that $\gamma$ must be 1 in the weak disorder isotropic limit, and asymptotically approach 0 in the large disorder limit.  

Relaxing the assumption about $p(\lambda,u,\upsilon)$'s exclusive dependence on $\lambda$ entirely, a more formal derivation of the Generalized DMPK equation was given in Ref.~[\onlinecite{mg}].  Going back to Eq.~(\ref{pint}), a model for the average over $M'$ known to reproduce the metal-insulator transition was used \cite{chalker}, and it was found that the integral over $d\mu(u)$ could be done exactly.  The resulting $p(\lambda'',\upsilon'')$ was Taylor expanded about $\lambda''=\lambda$, but it was assumed that an expansion of $\upsilon''$ about $\upsilon$ was unnecessary, especially in the insulating regime of interest because in that regime the eigenvectors settle into length-independent values.  The required integrals over $d\mu(\upsilon)$ were formally done using a mean field approxition using the following definitions:  
\be
K_{ab}\equiv \langle{k_{ab}}\rangle_{L_z} \equiv \sum_{\alpha} \langle |v_{a\alpha}|^2v_{b\alpha}|^2\rangle_{L_z},
\ee
and also
\be
\gamma_{ab} \equiv \frac{2K_{ab}}{K_{aa}}.
\ee
Here the angular bracket denotes an ensemble average. Requiring that the resulting equation obey probability conservation necessitated a renormalization of the Jacobian again so that $P(\lambda)$ was re-defined to be:
\be
\hat{P}(\lambda)= \prod_{a \neq b}^N|\lambda_a - \lambda_b|^{\gamma_{ab}} p(\lambda).
\ee
One could then write the resulting Generalized DMPK equation (GDMPK) as
\bea
\frac{\partial \hat{P}(\lambda)}{\partial t} &=&  
\sum_{a=1}^N K_{aa}\frac{\partial}{\partial \lambda_a}\lambda_a(1+\lambda_a) \cr
&\;&\times \left[\frac{\partial}{\partial \lambda_a}-\sum_{b \neq a} \frac{\gamma_{ab}}{\lambda_a-\lambda_b} \right] \hat{P}(\lambda).
\label{gendmpk}
\eea

As shown in Ref. [\onlinecite{mmw}], the matrix elements $K_{ab}$ can be explicitly evaluated numerically and contain information not only about dimensionality but also the critical point in 3d. The quasi 1d DMPK equation is recovered when $\gamma_{ab}\to 1$. In the above derivation the eigenvector correlations were neglected under the assumption that they become small at very large disorder, restricting the validity of Eq.~(\ref{gendmpk}) to the deeply insulating limit. Within this limit, the conductance distributions obtained analytically  from the  solutions of Eq.~(\ref{gendmpk}) agree well with numerical results obtained from tight binding Anderson model \cite{mmw,dm-10}.

Despite the successes, there are a few major criticism of the derivation of Eq.~(\ref{gendmpk}). The first one is that it is not clear if and/or under what conditions the eigenvector correlations can be neglected.  Second, it has heretofore been necessary to impose probability conservation as a separate constraint, which has required a `renormalization' of $J(\lambda)$, and redefinition of $P(\lambda)$ seemingly at odds with Eq. ~(\ref{Pdef}).  Last, there is the apparent conflict with the known constraint that in the limit  $|\lambda_a-\lambda_b| \to 0$, the level repulsion should be characterized by the symmetry parameter $\beta=1$ as determined by the exponent of the Jacobian and not by the matrix $\gamma_{ab}$, where $\gamma_{ab} \ll 1$ in the deeply insulating limit.  In this work we address all of these criticisms. To address the first and second criticisms we include  \textit{both} eigenvalue and eigenvector correlations systematically in a controlled way up to order $\delta L_z/l$.  We will find that including the additional eigenvector correlations results ultimately in the same GDMPK in Eq.~(\ref{gendmpk}), but also restores the Jacobian to its original form in Eq.~(\ref{Jac}), resulting in an equation which \textit{automatically} conserves probability.  Thus the results for $\mathcal{P}(g)$ obtained earlier in Refs.~[\onlinecite{mmw,dm-10}] turn out to be valid quite generally, beyond the deeply insulating limit. We address the last criticism by developing a self-consistent evolution equation for $K_{ab}$ and showing that $\gamma_{ab} \to 1$ as $|\lambda_a-\lambda_b| \to 0$, satisfying the symmetry requirement.  In addition, we obtain an exact relationship between the eigenvectors and the Lyapunov exponents $\nu_n$, defined by the relation $\langle \lambda_n \rangle \equiv \exp[(2L_z/l)\nu_n]$, leading to a sum rule for the Lyapunov exponents.

\section{The Generalized DMPK equation}

We will start with Eq.~(\ref{start}), and from there basically follow Ref.~[\onlinecite{mg}], except that we will not neglect the eigenvector correlations. For the sake of completeness and comparison, we will rederive some of the results already obtained in Ref.~[\onlinecite{mg}]. As in Ref.~[\onlinecite{mg}], we will go beyond the `isotropy' approximation of the original DMPK equation by keeping the $u$ and $v$ dependence of $p_{L_z}$ in addition to the $\lambda$ dependence. Using Eq.~(\ref{M}), and  equating $M''$ with $MM'^{-1}$, it then follows  that $u'' = u\cdot u''(\lambda, \lambda', vv'^{\dag})$. The invariance of the measure then allows one \cite{mg} to integrate over $u$, leading to an integral equation purely in terms of $p(\lambda,\upsilon)$,

\bea
p_{L_z+\delta L_z} (\lambda, v)&=&\int d\mu(u') d\mu(\lambda')d\mu(v') p_{\delta L_z}(\lambda', u',v')
\cr &\;& \;\;\;\; \times p_{L_z}(\lambda'', v'').
\eea
Since $p_{L_z+\delta L_z}(\lambda,\upsilon)$ is independent of $u$, it must be the case that $p_{\delta_{L_z}}(u',\upsilon',\lambda')$ is independent of $u'$.  So we set $u'=\upsilon'^\dagger$ for later convenience, and write
\bea
p_{L_z+\delta L_z} (\lambda, v)&=&\int d\mu(\lambda')d\mu(v') p_{\delta L_z}(\lambda', v')
\cr &\;& \;\;\;\; \times p_{L_z}(\lambda'', v'').
\eea

Next we write $\lambda'' = \lambda + \delta \lambda''$ and $v'' = v + \delta v''$ and Taylor expand $p(\lambda'',\upsilon'')$.  Thereafter we integrate both sides over $d\mu(\upsilon)$ and use integration by parts to write:
\bea
&&p_{L_z+\delta L_z} (\lambda)
= p_{L_z} (\lambda) \cr
&+&\sum_{a}\frac{\partial }{\partial \lambda_a}\int d\mu(v)p_{Lz}(\lambda, v)\langle \delta\lambda''_a\rangle_{M'}\cr
&+ & \frac{1}{2}\sum_{ab}  \frac{\partial^2 }{\partial \lambda_{a}\partial \lambda_b}\int d\mu(v) p_{L_z}(\lambda,v) \langle \delta \lambda''_{a}\delta \lambda''_b\rangle_{M'} \cr
&-&  \sum_{ab} \int d\mu(v) p_{L_z}(\lambda, v)\left \langle \frac{\partial  \delta v''_{ab}}{\partial v_{ab}} \right  \rangle_{M'} \cr
&-&  \sum_{a,bc}  \frac{\partial}{\partial \lambda_{a} } \int d\mu(v)   p_{L_z}(\lambda, v)\left \langle \frac{\partial \delta \lambda''_a \delta v''_{bc} }{\partial v_{bc}}\right \rangle_{M'}\cr 
&+&  \frac{1}{2}\sum_{ab,cd} \int d\mu(v)  p_{L_z}(\lambda, v) \left \langle \frac{\partial^2  \delta v''_{ab}\delta v''_{cd} }{\partial v_{ab}\partial v_{cd}} \right \rangle_{M'} \cr 
& + & \cdots 
\label{series} 
\eea

We now assume that the impurity averaged correlations $<\cdots >$ reach a limiting distribution that has a well-defined peak, possibly $\lambda$ - dependent, roughly coincident with their mean.  For example in the weak disorder limit, the correlations have a peak around the quasi-1d distribution. As checked numerically in Ref.~[\onlinecite{mmw}], this is always true provided the length $L_z$ is sufficiently large ($L_z \gg l$) (and the disorder is not so large that $\xi \sim l \sim a$ where $a$ is the lattice spacing). Note that we do not need to know what the peaked distribution actually looks like, it is enough to assume that such a distribution exists. In that case, equation (\ref{series}) can be rewritten as 
\bea
\frac{{\partial p_{L_z } (\lambda )}}{{\partial L_z }} &=& f(\lambda )p_{L_z } (\lambda ) + \sum\limits_a {g_a (\lambda )\frac{{\partial p_{L_z } (\lambda )}}{{\partial \lambda _a }}} \cr  &+& \sum\limits_{ab} {h_{ab} (\lambda )\frac{{\partial ^2 p_{L_z } (\lambda )}}{{\partial \lambda _a \partial \lambda _b }}}  + \cdots 
\label{pfgh}
\eea
where we define,
\bea
&& f(\lambda ) =  \frac{l}{{\delta L_{z}}} \left[ {\frac{1}{2}\sum\limits_{ab,cd} {\left\langle {\frac{{\partial ^2 \delta \upsilon ''_{ab} \delta \upsilon ''_{cd} }}{{\partial \upsilon _{ab} \partial \upsilon _{cd} }}} \right\rangle } 
 - \sum\limits_{ab} {\left\langle {\frac{{\partial \delta \upsilon ''_{ab} }}{{\partial \upsilon _{ab} }}} \right\rangle } }\right ]\cr
&&g_a (\lambda ) = \frac{l}{{\delta L_{z}}}\left[ {\left\langle {\delta \lambda ''_a \,} \right\rangle }  - \sum\limits_{bc} {\left\langle {\frac{{\partial \delta \lambda ''_a \,\delta \upsilon ''_{bc} }}{{\partial \upsilon _{bc} }}} \right\rangle  } \right] \cr
&&h_{ab} (\lambda ) = \frac{l}{{2\delta L_{z}}}\left\langle {\delta \lambda ''_a \,\delta \lambda ''_b \,} \right\rangle .
\label{fgh}
\eea
Here the angular brackets $\langle \cdots \rangle$  represent average over both $M'$ and $\upsilon$.
Note that in the limit $\delta L_z\to 0$, we only need to keep terms in the average that are at most linear in $\delta L_z/l$. In Ref.~[\onlinecite{mg}], only the terms corresponding to the changes in $\lambda''$ were assumed to be important, neglecting the changes in $\delta v''$.  Since these also contain terms linear in $\delta L_z/l$, a systematic expansion needs to keep all five terms. All other terms in the series expansion beyond the second derivative (represented by $(\cdots )$ in Eq.~(\ref{pfgh})) will be higher order in $\delta L_z/l$ and therefore need not be considered.  
In order to evaluate the changes $\delta\lambda''$ and $\delta v''$, we would like a matrix whose eigenvectors and eigenvalues are given by $\upsilon$ and $\lambda$ respectively.  Such a matrix, X, is given below \cite{mps}:     
\bea
X&=& \frac{1}{4}[M^{\dag}M + (M^{\dag}M)^{-1}-2I] \cr 
&=& 
\left(
\begin{array}{cc}
  \upsilon^{\dag}\lambda \upsilon  &   0   \\
   0 & \upsilon^{\dag}\lambda \upsilon     
\end{array}
\right).
\eea
A little algebra, using Eq.~(\ref{M}), is sufficient to demonstrate the last line.  From here we form $X''$ out of 
$M''=MM'^{-1}$ to calculate the perturbative corrections to $\lambda''$ and $\upsilon''$.  After some manipulations one obtains
\be
\upsilon ''^\dag  \lambda ''\upsilon ''\, = \upsilon ^\dag  \lambda \upsilon  + \upsilon ^\dag  W\upsilon 
\ee
where the matrix $W$ is given by (using $u'= \upsilon'^\dagger$ and expanding out to linear order in $\lambda'$)
\bea
W &=& \upsilon \upsilon '^\dag  \lambda '\upsilon '\upsilon ^\dag   + \frac{1}{2}\upsilon \upsilon '^\dag  \lambda '\upsilon '\upsilon ^\dag  \lambda  + \frac{1}{2}\lambda \upsilon \upsilon '^\dag  \lambda '\upsilon '\upsilon ^\dag  
\cr &-& \sqrt {\lambda (1 + \lambda )} \upsilon ^* \upsilon '^T \sqrt {\lambda '} \upsilon '\upsilon ^\dag 
\cr &-& \upsilon \upsilon '^\dag  \sqrt {\lambda '} \upsilon '^* \upsilon ^T \sqrt {\lambda (1 + \lambda )} \cr &+& \upsilon \upsilon '^\dag  \sqrt {\lambda '} \upsilon '^* \upsilon ^T \lambda \upsilon ^* \upsilon '^T \sqrt {\lambda '} \upsilon '\upsilon ^\dag .
\eea
The perturbative changes due to $M'$ are then given by 
\bea 
\delta\lambda''_n &=&W_{nn}+\sum_{i\ne n}\frac{|W_{in}|^2}{\lambda_n-\lambda_i}\cr
\delta v''^{\dag}_{mn} &=& \sum_{i\ne n}\frac{W_{in}}{\lambda_n-\lambda_i} v^{\dag}_{mi} -\frac{1}{2}\sum_{i\ne n}|W_{in}|^2  v^{\dag}_{mn}\cr
&+& \sum_{i\ne n}\sum_{j\ne n}\frac{W_{ij}}{\lambda_n-\lambda_i} \frac{W_{jn}}{\lambda_n-\lambda_j} v^{\dag}_{mi}\cr
&-& \sum_{i\ne n}\frac{W_{nn}}{\lambda_n-\lambda_i} \frac{W_{in}}{\lambda_n-\lambda_i} v^{\dag}_{mi}.
\label{perturbation}
\eea
Note that the eigenvector corrections so written are normalized to 2nd order.  To make further progress, we follow Ref.~[\onlinecite{mg}]  and use a very general model for the average over $M'$.  Since $\lambda'\to 0$ as $\delta L_z \to 0$, we use $\lambda' \propto \delta_{L_z}/l$  and the folowing `building blocks':

\begin{align}
&\langle v'_{ab}\rangle = \langle v'^*_{ab}\rangle = \langle v'_{ab}v'_{cd}\rangle = \langle v'^*_{ab}v'^*_{cd}\rangle = 0 \nonumber \\
&\sum_a^N\langle \lambda'_a v'^*_{a\alpha}v'_{a\beta}\rangle_{M'} =\frac{ \delta_{L_z}}{l} \delta_{\alpha\beta} \nonumber \\
&\sum_{ab} \langle \sqrt{ \lambda'_a\lambda'_b} v'^*_{a\alpha}v'^*_{a\beta}v'_{b\delta}v'_{b\gamma}\rangle_{M'} = \frac{ \delta_{L_z}}{l} \delta_{\alpha\beta}\delta_{\alpha\gamma}\delta_{\beta\delta} .
\label{model}
\end{align}
These are the building blocks used in the original DMPK equation and those suggested in Ref.~[\onlinecite{chalker}] to describe the metal-insulator transition. This allows us to build a \textit{systematic and controlled expansion} of $p_{L_z}$ in powers of the small parameter $\delta L_z/l$. All of the correlations in Eq.~(\ref{fgh}) can be written in terms of the following two results:

\bea
\langle{W_{mn}}\rangle_{M'}&=&\frac{\delta L_z}{l}(1+\lambda _n)\delta _{mn}\cr
&+&\frac{\delta L_z}{l}\sum\limits_{\alpha \beta}{\lambda_\alpha \upsilon_{m\beta}\upsilon_{n\beta}^*\left|{\upsilon_{\alpha\beta}}\right|^2}\notag\cr
\langle{W_{ij}W_{mn}}\rangle _{M'}&=& \frac{{\delta L_z }}{l}\Lambda_{in} \sum\limits_\alpha  {\upsilon_{i\alpha }^* \upsilon_{j\alpha }^* \upsilon _{m\alpha } \upsilon _{n\alpha }} \notag \cr
&+& \frac{\delta L_z }{l}\Lambda_{jm} \sum\limits_\alpha  {\upsilon _{m\alpha }^* \upsilon _{n\alpha }^* \upsilon_{i\alpha } \upsilon _{j\alpha}}
\label{Wcorrelations}
\eea
where we make the definition 
\bea
\Lambda_{ij}\equiv \sqrt{\lambda_i(1+\lambda_i)\lambda_j(1+\lambda_j)}.
\eea

As is also evident in Eq.~(\ref{fgh}), we need to take derivatives with respect to $\upsilon_{ab}$ and we use the following two basic equations for this,  
\bea
\frac{\partial v_{ab}}{\partial v_{ij}}= \delta_{ai}\delta_{bj}; \;\;\; 
\frac{\partial v^*_{ab}}{\partial v_{ij}}= -v^*_{aj}v^*_{ib}.
\eea
The latter is obtained by implicitly differentiating the relationship $\upsilon^\dagger\upsilon = 1$.  Finally, we need to then calculate averages over $\upsilon$.  To facilitate this, we make a random phase assumption, that averages over products of $\upsilon$'s will be zero unless their phases cancel.  With these in hand, and after lengthy algebra we find that $f$, $g$, and $h$ of Eq.~(\ref{fgh}) can all be written in terms of the $K$ matrix,

\bea
f({\bf{\lambda }}) &=& \sum\limits_{k \ne m} {\frac{{1 + 2\lambda _m }}{{\lambda _m  - \lambda _k }}\left[ {K_{mm}  - 2K_{mk} } \right]} \cr &+& \sum\limits_{k \ne m \ne n} {\frac{{\lambda _m (1 + \lambda _m )}}{{(\lambda _m  - \lambda _k )(\lambda _m  - \lambda _n )}}\left[ {K_{mm}  - 2K_{mk} } \right]} \cr
g_a ({\bf{\lambda }}) &=& (1 + 2\lambda _a )K_{aa}  + \sum\limits_{k \ne a} {\frac{{2\lambda _a (1 + \lambda _a )}}{{\lambda _a  - \lambda _k }}\left[ {K_{aa}  - K_{ak} } \right]} \cr
h_{ab} ({\bf{\lambda }}) &=& \lambda _a (1 + \lambda _a )K_{aa} \delta _{ab}.
\eea
Putting these results into Eq.~(\ref{pfgh}) and using the definition of $P(\lambda)$ from Eq.~(\ref{Pdef}) we find that we can write Eq.~(\ref{pfgh}) in full generality as  
\bea
\frac{\partial P(\lambda)}{\partial t}  &=&  
\sum_{a=1}^NK_{aa} \frac{\partial}{\partial \lambda_a}\lambda_a(1+\lambda_a) \cr
&\times & \left[
 \frac{\partial} {\partial \lambda_a} - \sum_{b\ne a} \frac{\gamma_{ab}}{\lambda_a-\lambda_b} \right] P(\lambda).
 \label{gdmpkP}
\eea
As one can see, this is the same Generalized DMPK equation as used in Ref.~[\onlinecite{mg}].  But this time we do not need to assume anything special about the eigenvector correlations, and indeed by including them, the equation now automatically conserves probability.  Finally, since this equation was derived under very general assumptions, we expect it to provide a description of the conductance distribution on both sides of and across the metal-insulator transition.  The only remaining problem is that Eq.~(\ref{gdmpkP}) implies that the level repulsion for neighboring levels is determined by the matrix $\gamma_{ab}$ which is not obviously related to the symmetry parameter $\beta$. In the insulating regime, it is known that on average $\gamma_{ab} \ll 1$ which seems to contradict this symmetry requirement.  In the next section we address this issue.  We show that under reasonable assumptions, $\gamma_{ab} \rightarrow \beta$ in the small separation limit.  To demonstrate this we develop an evolution equation for the matrix $K_{ab}$.

\section{Level repulsion between nearest neighbors}

We begin developing an evolution equation for $K_{mn}$ by expanding $K_{mn}$ out to linear order in $\delta L_z/l$.  We write
\bea
K_{mn}  + \delta K_{mn}  = \sum\limits_a {\langle\left| {\upsilon _{ma}  + \delta \upsilon _{ma} } \right|^2 \left| {\upsilon _{na}  + \delta \upsilon _{na} } \right|^2\rangle }
\eea
where the brackets denote an average over $\upsilon$. Then using the results of Eq.~(\ref{perturbation}), Eq.~(\ref{Wcorrelations}), and the random phase assumption we find that we can write the evolution of $K_{mn}$ with length $L_z$ as
\bea
\frac{dK_{mn}}{dt} &=& \sum\limits_{i \ne m} f_{im}(L_{ni}^{mi} - L_{nm}^{im}) + \sum\limits_{i \ne n} f_{in}(L_{mi}^{ni} - L_{mn}^{in}) \nonumber \\
&+& 2\sum\limits_{i \ne m} f_{im}L_{mi}^{mi}\delta_{mn} - 2f_{mn}L_{nm}^{nm}\delta_{m \ne n}
\label{Kevolution}
\eea
where $f_{ij}$ is given by,
\be
f_{ij} \equiv  \frac{\lambda_i(1 + \lambda_i) + \lambda_j(1 + \lambda_j)}{(\lambda_i - \lambda_j)^2}
\ee
and $L_{ik}^{jk}$ is given by,
\bea
L_{ik}^{jk}  &\equiv& \sum_{\alpha,\beta} \langle |\upsilon_{i\alpha}|^2|\upsilon_{k\alpha}|^2|\upsilon_{j\beta}|^2|\upsilon_{k\beta}|^2 \rangle \nonumber \\
&=& \langle k_{ik}k_{jk}\rangle .
\label{L}
\eea
The evolution equation for $K$ conserves probability for if we start with a $K_{mn}$ whose columns sum to 1, as it must, then one can show that the evolution equation preserves this property.  One can also see from equation Eq.~(\ref{Kevolution}) that the evolution of $K_{ab} \equiv \langle k_{ab} \rangle$ is determined purely by its second moment.  To gain some further insight we will split $L$ into the sum of two terms, its mean field approximation and fluctuations about the mean:
\be
L_{ik}^{jk}= \langle k_{ik}\rangle \langle k_{jk}\rangle + \Delta_{ik}^{jk}=K_{ik}K_{jk}+\Delta_{ik}^{jk}.
\label{meanfield}
\ee
Next we set $m$ = $n$ in Eq.(\ref{Kevolution}), and combine it with Eq.(\ref{meanfield}) to obtain:
\be
\frac{dK_{mm}}{dt} = K_{mm}^2\sum\limits_{i \ne m} f_{im}\left[\gamma_{mi}(1-\gamma_{mi}) + \frac{2\Delta _{mi}^{mi} - 4\Delta_{mm}^{mi}}{K_{mm}^2} \right].
\label{Kmm}
\ee
This equation has a few interesting consequences.  First, when fluctuations in $k$ are small, we may neglect the
$\Delta$ term.  Such fluctuations will be small compared to the mean value of $k$ in the weakly disordered
quasi-1d regime, and also in the 3d metallic regime, as shown in Ref.~[\onlinecite{mmw}]. Although these 
fluctuations are not generally small in the 3d strongly disordered regime, we may expect them to be negligible when the eigenvalues assume metallic-like configurations, namely, $1-\lambda_{n} / \lambda_{n+1} \ll  1/N$.   For such levels,   
Eq. (\ref{Kmm}) will reduce to:
\be
\frac{dK_{mm}}{dt} = K_{mm}^2\sum\limits_{i \ne m} f_{im}\gamma_{mi}(1-\gamma_{mi}).
\label{Kmmnodelta}
\ee
Now if we let nearest neighbors $\lambda_m$ and $\lambda_n$ approach each other in Eq.~(\ref{Kmmnodelta}), $f_{mn}$
will become singular.  But since $dK_{mm}/dt$ must be bounded, it must be that $\gamma_{mn}$ goes to 1.  
And so we see that in the small separation limit $|\lambda_n-\lambda_m|\to 0$, $\gamma_{mn}\rightarrow \beta=1$, as required. 
Thus for instance, even when $\lambda_1$ and $\lambda_2$ are far apart on average and consequently $\gamma_{12} \ll 1$ in the large disorder regime (as shown in numerical studies \cite{mmw}), in the limit where their separation goes to zero, the repulsion will be determined by the usual symmetry parameter equal to $\beta=1$ in the orthogonal case as considered here. 

\begin{figure}
\includegraphics[angle=0,width=0.45\textwidth]{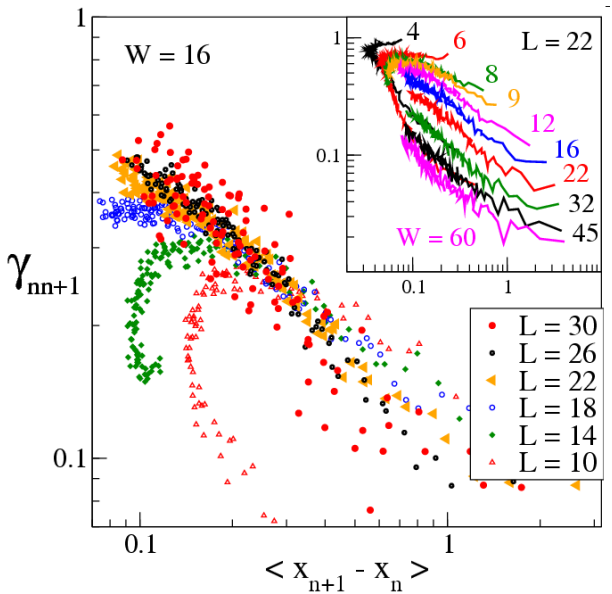}
\caption{Color online: $\gamma_{n,n+1}$ as a function of the difference $\langle  x_{n+1} - x_n\rangle $ for
statistical ensembles of cubic samples of different sizes $10 \le L \le 30$ for a given disorder strength $W=16$. (Only $N = 100$ samples are considered for $L=30$.) For comparison, the critical disorder \cite{mmw} $W_c\approx 9.4$. 
To avoid numerical inaccuracies, only channels with $\langle x_n - x_1\rangle  < 34$ were considered. The mean values of the logarithm of the conductance are $-4.07$ for $L=10$ and $-12.6$ for $L= 30$.
Typically, the level separation decreases with increasing $n$ up to $n=N/2$, and then start to increase again; this leads to two different `branches' (the upper one for $n < N/2$), as is visible for the two smallest sizes $L \le 14$. For larger $L$ this crossover occurs at larger $n$ and smaller separation, the lower branch eventually becoming numerically inaccessible for sufficiently large $L$. Inset:  The same ($\gamma_{n,n+1}$ vs the level separation) for a fixed length $L=22$ but for different values of $W$.}
\label{fig:gammaLW}
\end{figure}

Unfortunately it is difficult to access the above limit numerically since for large disorder the lowest levels have a statistically negligible probability to become close.  However it turns out that there are some levels near the middle of the spectrum that do come closer for any given disorder and for sufficiently large system size (see Fig. \ref{fig:gammaLW}).  Fig.~\ref{fig:gammaLW} illustrates the trend of how $\gamma_{n, n+1}$ changes when the separation of the levels becomes smaller. Here we have used the same anisotropic Anderson model as considered in Ref.~[\onlinecite{mmw}]. We consider a cubic system of size $L\times L\times L$ but with anisotropic nearest neighbor hopping elements  $t_z=1$ and $t_x=t_y=0.4$. The anisotropy guarantees that all channels are open, i.e. $N = L^2$. For a given disorder $W$ and a fixed system size $L$, the matrix $K$ and hence the parameters $\gamma_{mn}$ as well as the eigenvalues $\lambda_n$ are then evaluated using $10^4$ statistical samples within orthogonal symmetry. 

Since the eignevalues $\lambda_n=\cosh^{-2}(x_n/2)$ decrease exponentially with the size of the system, it turns out that many eigenvalues decrease below our numerical accuracy. 
Therefore, we consider only channels where $\langle x_n - x_1\rangle < 34$. This guarantees that 
$\lambda_n/\lambda_1 > 1.7\times 10^{-15}$, although at the same time the restriction does not enable us to consider all $\gamma_{mn}$ for a given $N$.  For instance, in Fig. \ref{fig:gammaLW}, $n\le 120$. Nevertheless, Fig.~\ref{fig:gammaLW} confirms our expectation that  $\gamma_{n,n+1}$ increases towards unity when the difference $ \langle x_{n+1}-x_n\rangle $ decreases. 

On the other hand, we expect Eq.~(\ref{Kmmnodelta}) to also describe a weakly disordered quasi-1d conductor.  In the large $L_z$ limit, the eigenvector correlations approach a limiting distribution.  In this case we must have
$dK_{mm}/dt = 0$.  And it follows that $\gamma_{mn} = 1$ as well, in accordance with the DMPK equation.  Indeed, if fluctuations are negligible we can solve for $K$ directly from Eq.~(\ref{Kevolution}).  Using Eq.~(\ref{meanfield}) with $\Delta = 0$, the evolution equation for $K$ reduces to:
\bea
\frac{dK_{mn}}{dt} &=& \sum\limits_{i \ne m} f_{im}(K_{ni}K_{mi} - K_{nm}K_{im})\nonumber \\
 &+& \sum\limits_{i \ne n} f_{in}(K_{mi}K_{ni} - K_{mn}K_{in}) \nonumber \\
&+& 2\sum\limits_{i \ne m} f_{im}K_{mi}^2\delta_{mn}\nonumber \\
&-& 2f_{mn}K_{nm}^2\delta_{m \ne n}
\label{Kevolution2}
\eea
It is straightforward to show that the $L_z$ independent solution to this equation is
\bea
K_{mn} = \frac{1+\delta_{mn}}{N+1} 
\eea
as required by the quasi 1d DMPK equation. In order to move beyond the small fluctuation limit, and model more generally a strongly disordered insulator, especially near the critical point, it is expected that $\Delta$ will become important. Though little can be said at this point, one implication of Eq.~(\ref{Kmm}) is that if $K_{mm}$ reaches a limiting distribution in $L_z \gg L$ strongly disordered limit, the steady state value of $\gamma_{mn}$ will depend on these fluctuations.  

\section{The K-matrix and the Lyapunov exponents}
\begin{figure}
\includegraphics[angle=0,width=0.45\textwidth]{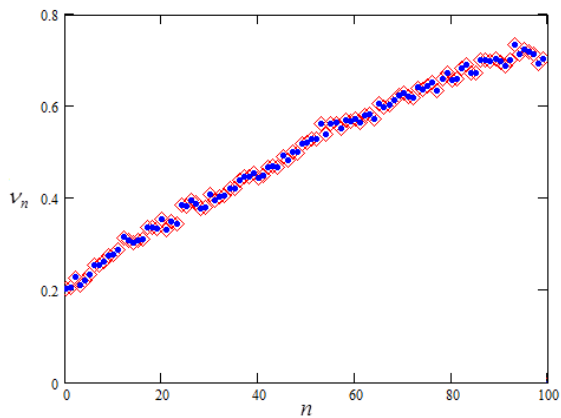}
\caption{(Color online) The Lyapunov exponents $\nu_n$ plotted vs. $n$. The red diamonds are from numerical simulations ($L = 10$, $W = 20$, and $<$ln g$>$ = -7.1); the solid blue circles are from Eq.~(\ref{nu}).}
\label{fig:nus}
\end{figure}

Finally, we develop an equation relating the Lyapunov exponents and the matrix K.  The Lyapunov exponent $\nu$ is defined by 
\bea
\lambda (L_z  \to \infty ) = e^{2\nu L_z }.
\eea
In order to develop an equation for $\nu$ we write
\bea
\lambda (L_z  + \delta L_z ) - \lambda (L_z ) = e^{2\nu (L_z  + \delta L_z )}  - e^{2\nu L_z }.
\eea
Solving for $\nu$ we have
\bea
\nu  = \frac{1}{{2\delta L_z }}\ln \left( {1 + \frac{{\delta \lambda}}{{\lambda}}} \right).
\eea
Expanding out to linear order in $\delta L_z/l$ and averaging over disorder we obtain
\be
\nu_n=\frac{l}{2\delta L_z}\left[\left \langle \frac{\delta\lambda_n}{\lambda_n}\right \rangle_\upsilon -\frac{1}{2}\left \langle \left(\frac{\delta\lambda_n}{\lambda_n}\right)^2 \right\rangle_\upsilon \right].
\ee
Using Eqs.(\ref{perturbation}) and (\ref{Wcorrelations}), we finally arrive at
\be
\nu_n=\frac{1}{2}K_{nn} + \sum_{m \ne n} \frac{1+\lambda_n}{\lambda_n-\lambda_m}K_{nm}
\label{nu}
\ee
for all disorder. A similar equation was obtained in Ref.~[\onlinecite{chalker}]. Eq.~(\ref{nu}) has been verified numerically in the insulating regime, where most doubt concerning its validity would reside.  When plots of $\nu_n$ obtained via numerical simulations using tight binding Anderson model and from Eq.~(\ref{nu}) are overlaid, as shown in Figure \ref{fig:nus}, the points are indistinguishable.  

Interestingly, if we sum this equation over $n$, we obtain the following relationship, independent of $K$ and $\lambda$,
\be
\sum_{n=1}^N \nu_n = \frac{N}{2}.
\ee
This relation has been known to be true in the diffusive limit \cite{pichard}, but the present calculation implies it is true for large disorder as well.   This has been confirmed numerically.

Note that these relations are independent of any model for the matrix $K$. A $2$-parameter model for $K$ was chosen in Ref.~[\onlinecite{mmw}] for large disorder, based entirely on numerical studies. The exact relationship should allow us to build a better model based on the known properties of the Lyapunov exponents. 

\section{Summary and Conclusion}

In this work we developed a systematic and controlled derivation of the Generalized DMPK equation by expanding the probability distribution $p_{L_z+\delta L_z}(\lambda)$ in a power series $\delta L_z/l$ and keeping all terms linear in $\delta L_z/l$ in the limit $\delta L_z/l \to 0$. This means including certain eigenvector correlations that had been neglected before.  The additional correlations do not change the Generalized DMPK equation obtained earlier; instead, their inclusion automatically conserve probability by allowing the evolution of the Jacobian without having to redefine it with a disorder dependent exponent. In addition to providing a broader applicability of the Generalized DMPK equation such as the conductance distributions obtained in Ref.~[\onlinecite{dm-10}], the derivation shows how the repulsion of closely spaced neighboring eigenvalues remain consistent  with symmetry requirements, even though it can be very different when the eigenvalues are far apart. Moreover, we obtain an exact result relating the correlation matrix $K$ and the eigenvalues $\lambda$ with the Lyapunov exponents $\nu$ as well as a sum rule for the exponents independent of  $K$ or $\lambda$.

The challenge remains to construct a reasonable model for the phenomenological matrix $K$. The evolution of $K$ involves quantities that can not be factored into products of $K$'s in general, and at this point an analytic solution seems too complicated. On the other hand a crude phenomenological $2$-parameter model suggested by numerical studies and used in Refs.~\onlinecite{dm-10,mmw} seems to work quite well in the strongly disordered regime, although further numerical studies are needed to check if/how the model might change when e.g. approaching the metal-insulator critical point. We hope that the relationships with the Lyapunov exponents obtained here should provide useful insights.  

PM acknowledges financial support from  the Slovak Research and Development	Agency under the contract No. APVV-0108-11.

\end{document}